\documentclass[a4paper,fleqn,usenatbib]{mnras}

\usepackage{newtxtext,newtxmath}

\usepackage[T1]{fontenc}
\usepackage{ae,aecompl}

\usepackage{graphicx}	
\usepackage{amsmath}	

\usepackage{amssymb}	

\title[Classical bulges in S0 galaxies]{Why are classical bulges more common in S0 galaxies than in spiral galaxies?}

\author[P. K. Mishra et al.]{
Preetish K. Mishra,$^{1}$\thanks{E-mail: preetish@ncra.tifr.res.in}
Yogesh Wadadekar,$^{1}$\thanks{E-mail: yogesh@ncra.tifr.res.in}
Sudhanshu Barway,$^{2}$\thanks{E-mail: sudhanshu.barway@iiap.res.in  }
\\
$^{1}$National Centre for Radio Astrophysics, TIFR, Post Bag 3, Ganeshkhind, Pune 411007, India\\
$^{2}$Indian Institute of Astrophysics (IIA), II Block, Koramangala,  Bengaluru 560034, India
}
\date{Accepted for publication in MNRAS}

\pubyear{2017}

\begin{document}
\label{firstpage}
\pagerange{\pageref{firstpage}--\pageref{lastpage}}
\maketitle

\begin{abstract}
 
 In this paper, we try to understand why the classical bulge fraction
 observed in S0 galaxies is significantly higher than that in spiral
 galaxies. We carry out a comparative study of the bulge and global
 properties of a sample of spiral and S0 galaxies in a fixed
 environment. Our sample is flux limited and contains 262 spiral and
 155 S0 galaxies drawn from the Sloan Digital Sky Survey. We have
 classified bulges into classical and pseudobulge categories based on
 their position on the Kormendy diagram. Dividing our sample into bins
 of galaxy stellar mass, we find that the fraction of S0 galaxies
 hosting a classical bulge is significantly higher than the classical
 bulge fraction seen in spirals even at fixed stellar mass. We have
 compared the bulge and the global properties of spirals and S0
 galaxies in our sample and find indications that spiral galaxies
 which host a classical bulge, preferentially get converted into S0
 population as compared to pseudobulge hosting spirals. By studying
 the star formation properties of our galaxies in the $NUV-r$
 color-mass diagram, we find that the pseudobulge hosting spirals are
 mostly star forming while the majority of classical bulge host spirals
 are in the green valley or in the passive sequence. We suggest that
 some internal process, such as AGN feedback or morphological quenching due to the massive bulge, quenches these classical bulge hosting spirals and transforms them into S0 galaxies, thus resulting in the observed predominance of
 the classical bulge in S0 galaxies.
 
\end{abstract}




\section{Introduction}
\label{sec:intro}

S0 galaxies are a class of galaxies with absent (or very faint) spiral
arms in their disc. Traditionally, these galaxies are regarded as an
intermediate transition class between two other major morphological
classes, namely ellipticals and spirals, on the Hubble tuning fork
diagram \citep{Hubble1936}. The formation and evolution of S0 galaxies
is a field of active research and a major effort has been carried out
to understand their nature. S0 galaxies are considered as transformed
spiral galaxies \citep{Barway2009} and their formation scenarios can
be broadly classified into two categories. The first category
  of morphological transformation is the one driven by mergers or
  other gravitational interactions with neighbouring galaxies. A
  spiral galaxy undergoing a major merger with a companion of similar
  mass or undergoing a series of minor mergers with smaller (with mass
  ratio 1:7 or more) companions can transform into an S0 galaxy
  (\cite{Querejeta2015,Tapia2017} and references therein). Tidal
  interaction and harassment of a spiral galaxy residing in a dense
  environment with its neighbouring galaxies can also lead to
  disappearance of spiral arms in the disc \citep{Moore1996,
    Milhos2003}.  The second category of formation scenario explains
the formation of S0 galaxies via quenching of star formation in the
progenitor spiral galaxy. This quenching can happen either due to
environmental processes or by some internal mechanism. Environmental
quenching of star formation can take place due to processes like ram
pressure stripping, tidal interaction and halo quenching or starvation
\citep{Gunn1972,Moore1996,Larson1980,Peng2015} which are
characteristic of galaxy clusters and large groups having a massive
dark matter halo, while internal quenching can be due to AGN feedback
or stability of disc against star formation due to a massive bulge
etc. \citep{Cox2006,Martig2009}. All of the above mentioned
  processes are the potential ways to transform a spiral galaxy into
  an S0 galaxy via disc fading and disappearance of spiral
  arms. \citep{Bekki2002,Barway2007,Barway2009,Bekki2011}. The
dominance of one formation scenario over anoother depends on the mass
of the progenitor spiral and the environment in which it resides but
it is not easy to distinguish between these scenarios using
observations. \par

If the disc morphology differentiates the S0 and spirals into separate
morphological classes, the central component of the galaxy --the
bulge--, connects them. It is known that bulges of disc galaxies can
be broadly classified in two classes: classical bulges and
pseudobulges \citep{Kormendy2004}. Classical bulges are formed by fast
and violent processes such as major mergers or by sinking and
coalescence of giant gas clumps found in high redshift discs
\citep{Elmegreen2008, Kormendy2016}. They are kinematically hot,
featureless spheroids following the same scaling relations on the
fundamental plane \citep{Djorgovski1987} as elliptical
galaxies. Pseudobulges are thought to be formed by the slow
rearrangement of gaseous material from the disc to the central region
of galaxies, driven by non-axisymmetric structures such as bars, ovals
etc., or via minor mergers \citep{Eliche-Moral2011,
  Kormendy2004}. Pseudobulges are rotationally supported systems
having discy morphology and mixed stellar populations. Even though the
two types of bulges differ in their properties from one another, one
finds that both the types are hosted by spirals and S0
galaxies. However, the interesting fact is that while the frequency of
occurrence of classical and pseudobulges is comparable in spiral
galaxies, S0 galaxies prefer to host a classical bulge. The fraction
of S0 galaxies which host a pseudobulge is a rather low value which
ranges from 7.5\% to 14\% as quoted in literature, while for spirals
this value ranges from 32\% to 42.5\%
\citep{Gadotti2009,Vaghmare2013,Mishra2017}. \par

If S0 galaxies are transformed spirals, then naively one expects the
classical bulge (or pseudobulge) fraction to be roughly equal in both
spirals and S0 galaxies, unless the process which makes an S0 galaxy
out of a spiral galaxy also changes the bulge type. Then it becomes interesting to ask the question
: why do S0 galaxies prefer to host a classical bulge? Is it because the
process which forms majority of S0s out of spirals changes the bulge
type or is it due to preferential conversion of classical bulge host
spirals into S0 galaxies?

One must, however, be watchful of, the observational biases that seep
in while comparing the bulge fraction in the two morphological
classes. Previous studies on galaxy bulges have shown that classical
bulges are more commonly found in high mass galaxies and in galaxies
residing in a denser environment \citep{Fisher2011,Kormendy2016,Mishra2017a}. This might introduce a bias in the comparison of classical bulge fraction in spiral and S0 galaxy population because S0
galaxies are more commonly found in high density environments and, are
on an average more massive compared to spirals. It is also known that
galaxy stellar mass and environment are correlated and one finds more
massive galaxies in high density environments. A higher mass spiral,
which is more likely to host a classical bulge, might get transformed
into an S0 galaxy due to quenching of star formation due to
environmental processes such as ram pressure stripping, starvation,
tidal interaction etc. Therefore, for a comparison of classical bulge
fraction in these two morphological classes one must account for 
this environmental bias or, at the very least, minimise this bias while choosing the galaxy sample.

In this work, we have explored the possible reason for the observed mismatch between classical bulge fraction in spirals and S0 galaxies. Forming a sample of spirals and S0 galaxies in a fixed environment, we compare the global and bulge properties of these two morphological classes in fixed bins of mass. The organization of this paper is as follows, Section \ref{sec:data} describes our data and sample selection. In Section \ref{sec:result} we present our results which are discussed in Section \ref{sec:discussion} before we present the summary of findings and interpretations in Section \ref{sec:sum}. Throughout this work, we have used the WMAP9 cosmological parameters: $H_0$ = 69.3 km s$^{-1}$Mpc$^{-1}$,$\Omega_m$= 0.287 and $\Omega_{\Lambda}$= 0.713

\begin{figure*}

\includegraphics[width=.34\textwidth]{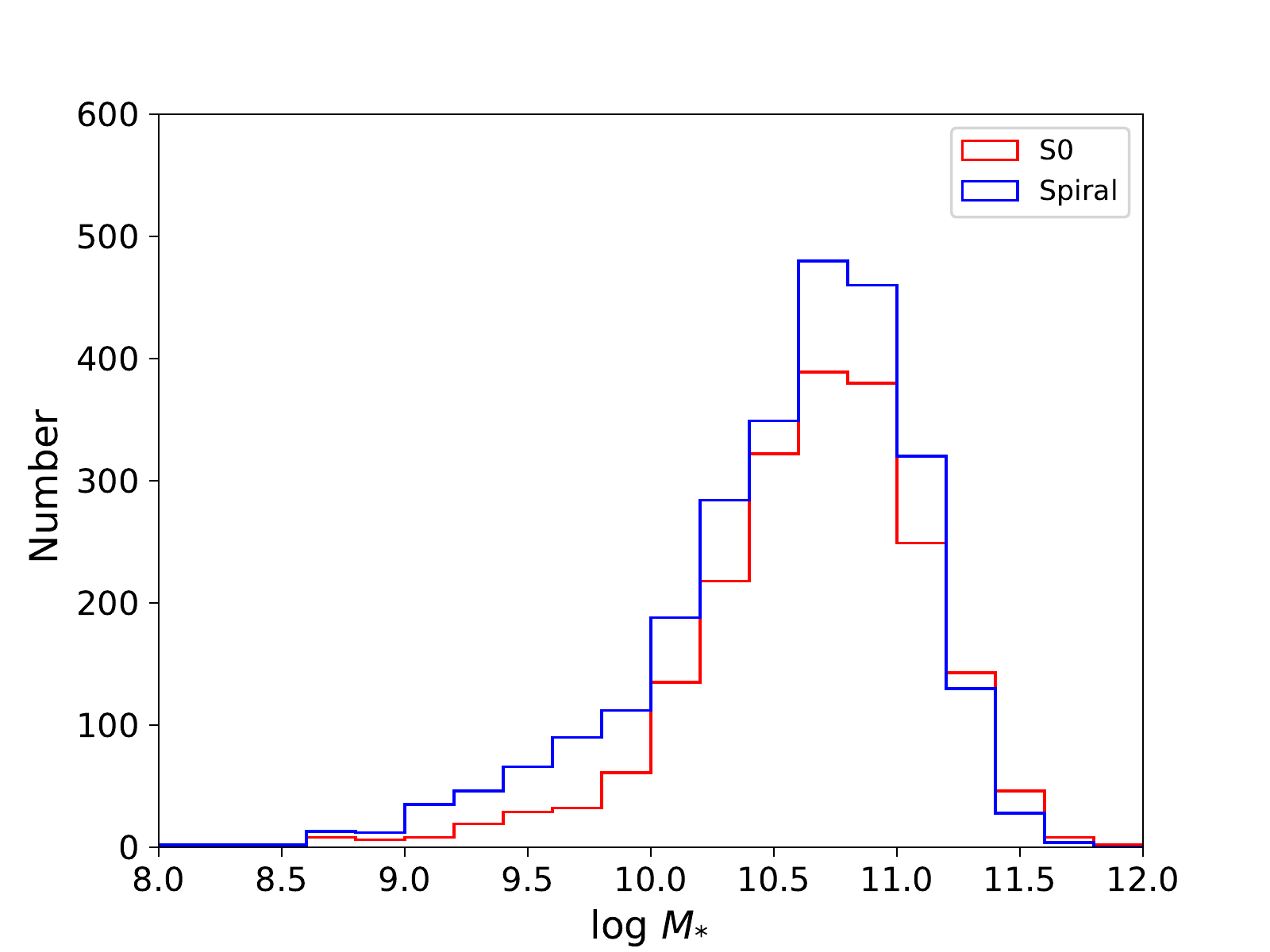}\hspace{-1.5em}
\includegraphics[width=.34\textwidth]{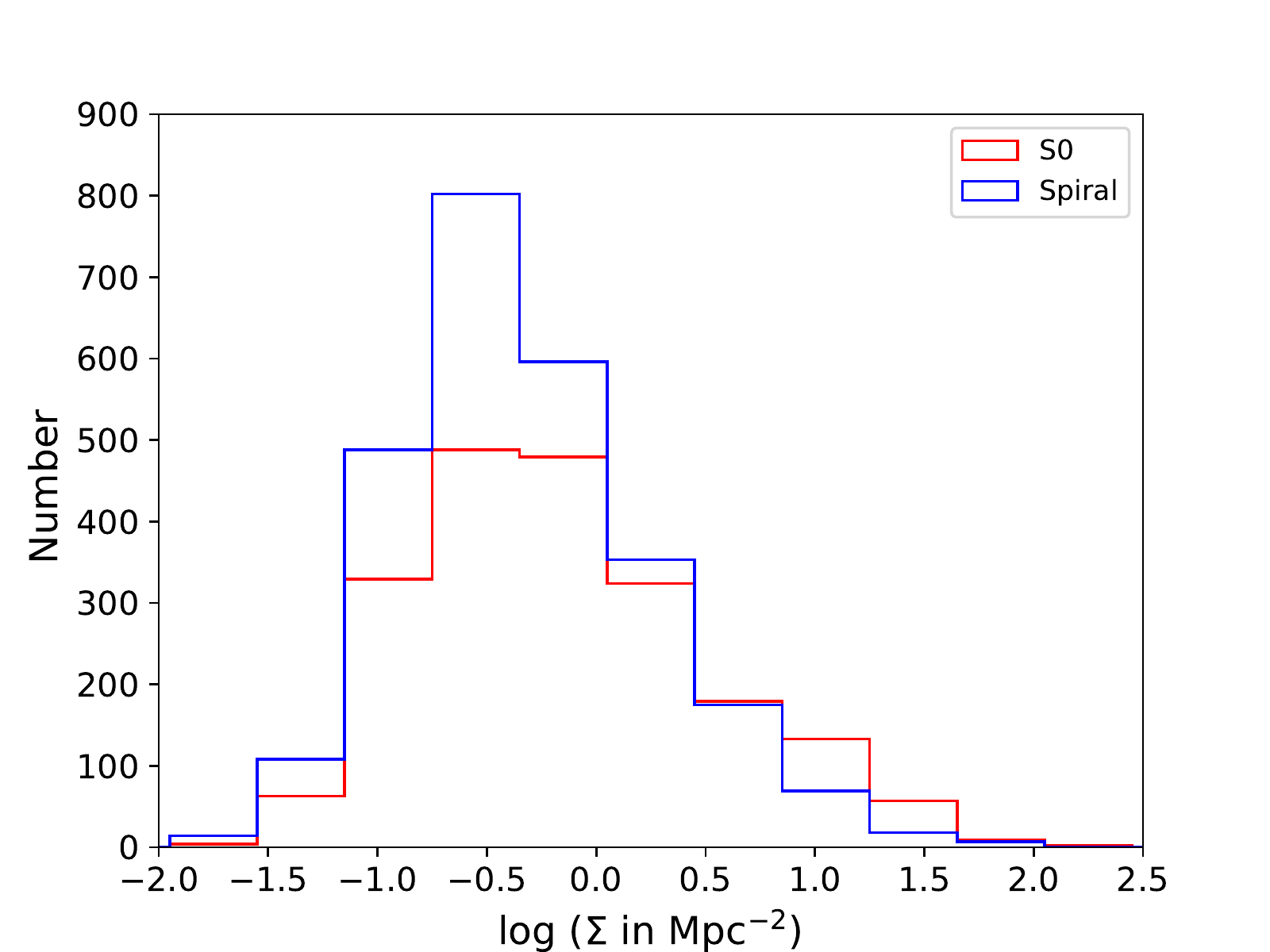}\hspace{-1.5em}
\includegraphics[width=.34\textwidth]{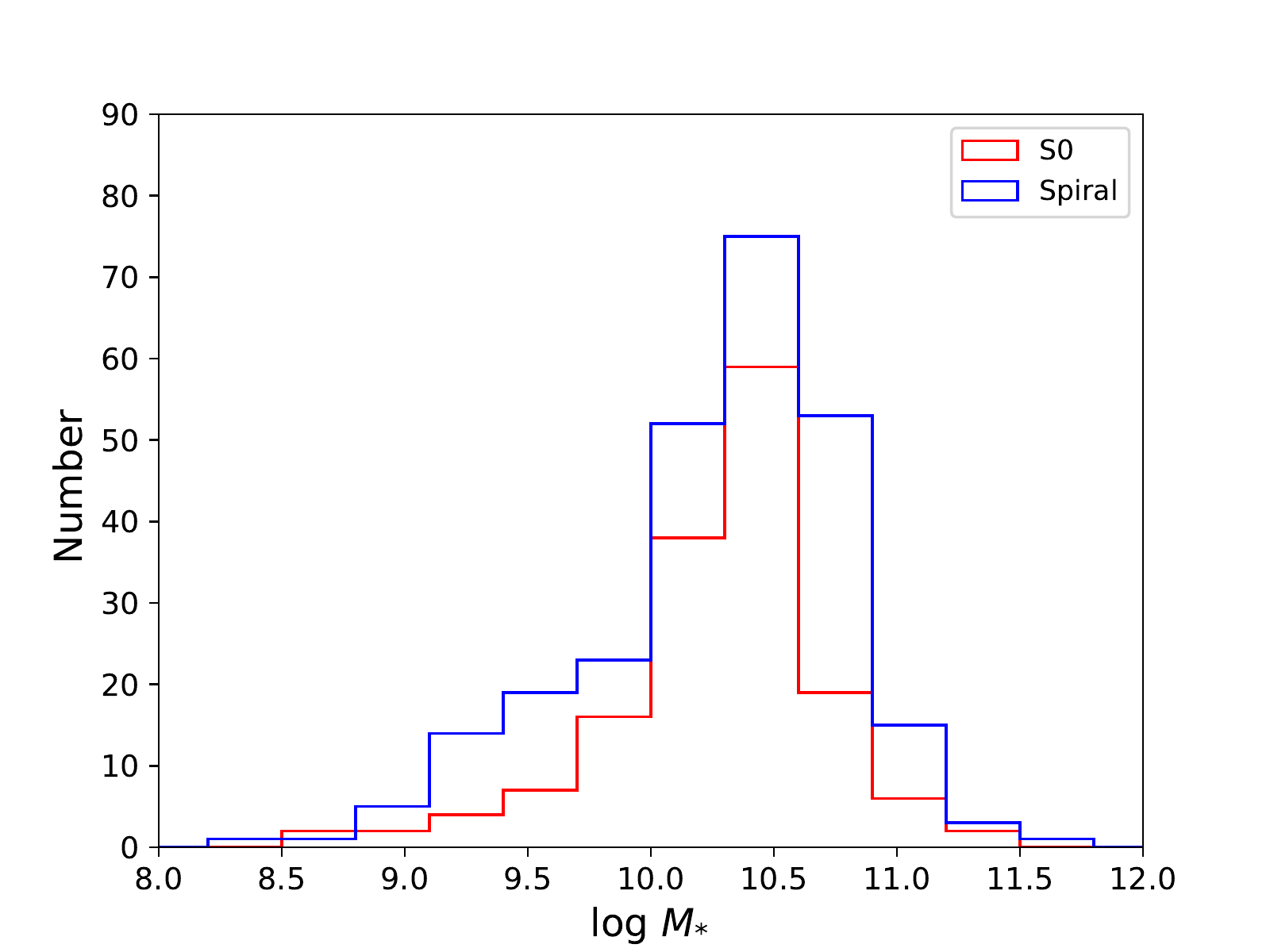}

\caption{The left and middle panels show the distribution of stellar mass (in units of log($M_{\odot}$)) and environment for the spiral and S0 galaxies in parent sample. The right panel shows the stellar mass distribution of spirals and S0 galaxies after application of environmental ($-0.75 <$log$\Sigma< -0.35$) and halo mass ($<10^{12} M_{\odot}$) cuts. }
\label{fig:fig1}

\end{figure*}

\section{Data and sample selection}
\label{sec:data}

In order to construct a statistically significant sample for our
study, we start with data provided in the \cite{Nair2010} catalogue
which is a catalogue of visual morphological classification for about
14,000 spectroscopically targeted galaxies in the SDSS DR4. The
\cite{Nair2010} catalogue is a flux limited sample with an extinction
corrected limit of $g < 16$ mag in the SDSS $g$ band, spanning the
redshift range $0.01 < z < 0.1$. In addition to {\bf visual}
morphology information for each galaxy given in the form of the Hubble
T type, the catalogue also lists other relevant quantities for each
object such as stellar mass taken from \cite{Kauffmann2003}, fifth
nearest neighbour environmental density from \cite{Baldry2006}, group
membership information and host dark matter halo mass taken from
\cite{Yang2007}. To obtain information of structural parameters of
these galaxies, we cross matched the \cite{Nair2010} catalogue with
data provided in \cite{Simard2011} catalogue. \cite{Simard2011} is a
catalogue of two dimensional, point-spread-function-convolved,
bulge+disc decompositions in the $g$ and $r$ bands for a sample of
1,123,718 galaxies from the SDSS DR 7. The cross match resulted in
12,063 galaxies.\par

\cite{Simard2011} fit each galaxy with three different models of the
light profile : a pure S\'ersic model, an $n_b$ = 4 bulge + disc
model, and a S\'ersic (free $n_b$) bulge + disc model. They also
provide an F-test probability criterion based on which one can choose
the most appropriate model of light profile for a given galaxy. Since
we are interested in studying bulges of disc galaxies, we have chosen
only those galaxies where a bulge+disc model is preferred over a
single S\'ersic model. For the elliptical galaxies in our sample, we
chose to use $n_b$ = 4 bulge + disc model as we have found that
majority of ellipticals in our sample are better fitted by this model
as compared to the S\'ersic (free $n_b$) bulge + disc model. We chose
the $n_b$ bulge + disc model for the disc galaxies in our sample as it
is known from the literature \citep{Graham2001,Balcells2003,
  Barway2009} that bulges of S0s and spirals span a wide range of
values of the S\'ersic index.\par

After making an appropriate choice of light profile model for each
galaxy in our sample, we retain only those galaxies for which we
have a reliable estimate of bulge S\'ersic index ($n_b$). We do not
discuss the reliability of $n_b$ estimate in this paper. This has been
discussed in our previous work \citep{Mishra2017} and we adopt the
same criteria to exclude unreliable fits. Interested readers are
referred to \cite{Mishra2017} for details. 

We have identified galaxies which host a bar using the flag provided
in \cite{Nair2010}, and have removed them from our sample as
\cite{Simard2011} does not take bar into account in their fits and
this might cause significant error in the estimation of bulge
parameters. Application of these selection cuts on the initial sample of 
12,063 galaxies leaves us with a sample of 1742 ellipticals and 4697
disc galaxies which we refer to as the parent sample in this
paper. \par

For a comparative study of bulge fraction in spiral and S0 galaxies,
one must account for the previously mentioned environmental bias in
the sample. In order to do this, we chose to restrict our sample in a
sufficiently narrow and low density regime of environmental parameter
space. The distribution of stellar mass and environmental density
parameter ($\Sigma$) measuring the fifth nearest neighbour density for
the galaxies in our parent sample are shown in left and middle panels
of Figure \ref{fig:fig1} respectively. We restricted our sample of
spirals and S0 galaxies in the bin of log $\Sigma$ where the
distribution peaks. This bin has range $-0.75 <$log $\Sigma< -0.35$
and corresponds to an intermediate to low local density
environment. We further impose a cut based on the dark matter halo
mass to which a certain galaxy belongs. We have taken only those
galaxies where the mass of the host dark matter halo is less than
$10^{12} M_{\odot}$. This is the halo mass limit above which mechanism
of quenching of star formation due to starvation dominates
\citep{Dekel2006}. Application of these two cuts on environment leaves
us with a sample of 170 S0 and 353 spiral galaxies.  We refer to this
sample as the reduced sample. This sample consists mainly of field
galaxies or galaxies residing in small groups, the largest of which has
7 members. For the reduced sample, environment has little
influence.\par

After accounting for the environmental bias, we plan to compare the
bulge fraction in S0 and spiral galaxy population in fixed bins of
stellar mass. We also want to see if the same class of bulges hosted
by spiral and S0 galaxies are different in some of their properties
which might help us to explain the observed mismatch in the bulge
fraction. We have the available measurements of central velocity
dispersion, probing the kinematics of central region of galaxies, from
\cite{Nair2010} catalogue. We also obtained information on the central
stellar population parameters like the $D_n(4000)$ index (as
defined in \cite{Balogh1999}) from the table {\it galSpecIndx} using the
SDSS DR13. All of these measurements are derived from galaxy spectra
coming from the central 3 arcsecond diamter of each object as probed by the SDSS
fibre aperture. We wanted these values to be representative of bulge
region of galaxies. But the galaxies in our sample have different
sizes and are distributed across the redshift range $0.01 < z < 0.1$ and
therefore there is a chance that bulge light coming from the central 3
arcsecond is contaminated significantly by the disc light. In
order to minimise this contamination, we found the radius at which
disc light starts to dominate the bulge light by plotting light
profiles of disc and the bulge from \cite{Simard2011}
decompositions. We demanded that for all the galaxies in the reduced
sample this radius should be greater than or equal to the radius of the SDSS
fiber aperture. Application of this criterion gave us our final sample
of 155 S0 and 262 spiral galaxies. The mass distribution of the final
sample is plotted in the right panel of Figure \ref{fig:fig1}.

\begin{table*}
	\centering
	\caption{Number of classical and pseudobulge hosting spirals and S0 in different stellar mass bins. }
	\label{tab:table}
	\begin{tabular}{lcccccc} 
		\hline
		Mass range (log$(M_*/M_{\odot})$) &8.2 - 8.8& 8.8 - 9.4& 9.4 - 10.0& 10.0 - 10.6&10.6 - 11.2&11.2 - 11.8 \\
		\hline
		
		No. of spiral galaxies & 2 & 19 &42 &127 & 68 & 4\\
		Pseudobulge host spirals & 2 & 18 &36 &71 & 25 & 1\\
		Classical bulge host spirals & 0 & 1 &6 &56 & 43 & 3\\
		Spiral classical bulge fraction (\%)& 0 & 5.3 &14.3 &44.1 & 63.2 & 75\\
		\hline
		No. of S0 galaxies & 2 & 6 &23 &97 & 25 & 2\\
		Pseudobulge host S0s & 2 & 5 &10 &8 & 1 & 0\\
		Classical bulge host S0s & 0 & 1 &13 &89 & 24 & 2\\
		S0 classical bulge fraction (\%)& 0 & 16.7 &56.5 &91.7 & 96.0 & 100\\		
		\hline
	\end{tabular}
\end{table*}
 
\section{Results}
\label{sec:result}
\subsection{Bulge classification}

We have classified bulges into classical and pseudobulge types based
on their position on the Kormendy diagram \citep{Kormendy1977}. This
diagram is a plot of the average surface brightness of the bulge
within its effective radius <$\mu_e$> against the logarithm of the
bulge effective radius $r_e$. This bulge classification scheme which
makes use of the Kormendy diagram was proposed by
\cite{Gadotti2009}. Elliptical galaxies are known to obey a tight
linear relation on this diagram. Classical bulges being structurally
similar to elliptical galaxies follow the same scaling
relation. Pseudobulges, on the other hand are structurally different
from ellipticals and hence they lie away from this
relation. \cite{Gadotti2009} has proposed that bulges which deviate
more that three times the r.m.s. scatter from the best fit relation
for ellipticals be classified as pseudobulges while those falling
within this scatter be classified as classical bulges. This physically
motivated bulge classification has also been used in recent works
\citep{Vaghmare2013,Neumann2017,Mishra2017} \par

The Kormendy relation for our sample was obtained by fitting ellipticals
in our parent sample using $r$ band data. The equation
for the best fit line is \\

$\langle\mu_b (< r_e)\rangle$ = $(2.330 \pm 0.047)$ log($r_e$) + $(18.160 \pm 0.024)$
\\

The rms scatter in $\langle\mu_b (< r_e)\rangle$ around the best fit
line is 0.429. All galaxies which lie away more than 3 sigma scatter
from this relation are classified as pseudobulge hosts while those
within this scatter are classified as classical bulge hosts. \par

After application of this criterion, we find that out of 262 spiral
galaxies in our final sample the number of classical and pseudobulge
hosts are 109 (41.6\%) and 153 (58.4 \%) respectively. For the total
155 S0 galaxies, we find 129 (83.2\%) of them host a classical bulge
while the rest 26 (16.8\%) are pseudobulge hosts. One can already
notice the large mismatch between the fraction of classical bulge host
galaxies in spiral and S0 morphology class even in fixed
environment. We now divide the mass range of our final sample in 6
different stellar mass bins. The stellar mass bin divisions are
log$(M_*/M_{\odot})$ = [8.2, 8.8, 9.4, 10.0, 10.6, 11.2, 11.8], and
Table \ref{tab:table} contains the relevant number and fraction of
classical and pseudobulge hosting spiral and S0 galaxies in each bin
of stellar mass. One can notice from Table \ref{tab:table} that
fraction of galaxies which host a classical bulge increases with
increase in mass for both the galaxy morphologies. But the interesting
thing to note is that in all the mass bins which have enough galaxies
for a statistically meaningful comparison, the classical bulge
fraction in S0 galaxies is significantly higher as compared to the
spirals. This confirms the notion that the observed mismatch between
classical bulge fraction in spiral and S0 galaxies is not driven by
any environmental or mass biases of the sample, but is an
observational fact.  \par

We now return to our original question which is to understand what makes the classical bulge fraction so different in spirals and S0 galaxies. The possible reasons include pseudobulges getting transformed into classical bulges as spirals get converted into S0s or maybe somehow only the classical bulge hosting spirals are preferentially getting converted to S0 population. But before settling on any of these two reasons, we first try to find further clues to the problem by comparing the bulge and the global properties of spiral and S0 galaxies. Such a comparison might help us to identify the possible progenitors of classical bulge hosting S0 galaxies and to understand the reason of the observed difference between classical bulge fraction in spirals and S0 galaxies.  \par

\subsection{Bulge and global properties of S0 and spiral galaxies}

\begin{figure*}
	\includegraphics[width=0.75\textwidth]{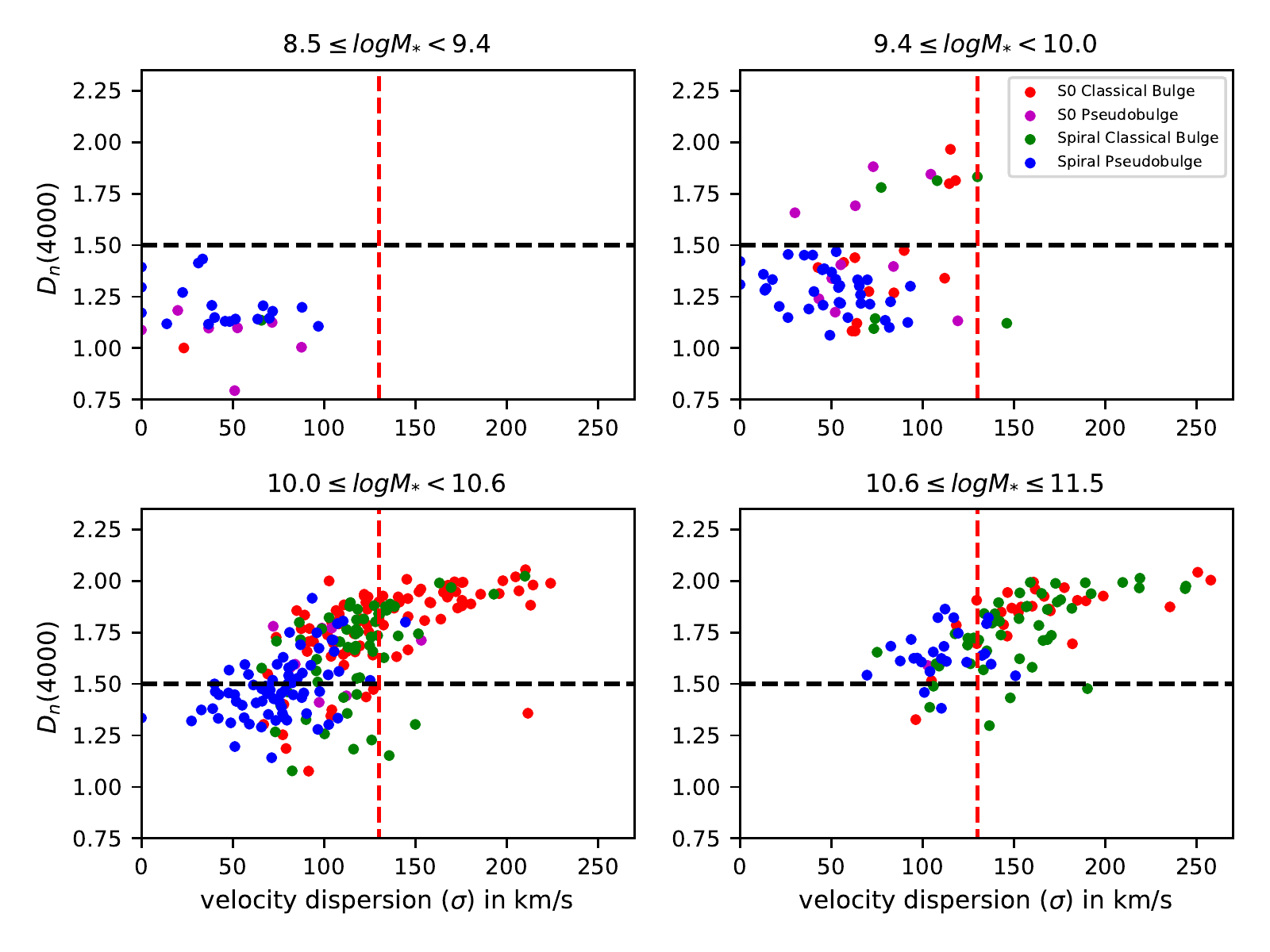}
   \caption{Plots showing the position of classical and pseudobulge hosting spiral and S0 galaxies on  the $D_n(4000)$-central velocity dispersion ($\sigma$) plane. The stellar mass range of galaxies in each panel is stated on the top of each panel in units of log($M_{\odot}$). The black dashed line $D_n(4000)= 1.5$ separates bulges into young (with $D_n(4000)<1.5$ ) and old ($D_n(4000)\geq1.5$) population. The red line at $\sigma = 130$ km/s has been put for reliability check of pseudobulge classification. Pseudobulges falling to the right to this line are unreliable. The plot shows that higher mass galaxies have older and kinematically hot bulges. While the population of classical and pseudobulges are expectedly different in their properties, the classical bulges hosted by spirals and S0 have sufficient overlap in the $D_n(4000)- \sigma$ space pointing to a similarity in properties.}
   \label{fig:Dsigma}
\end{figure*}

\begin{figure*}
  \centering
  \begin{minipage}[b]{0.49\textwidth}
    \includegraphics[width=\textwidth]{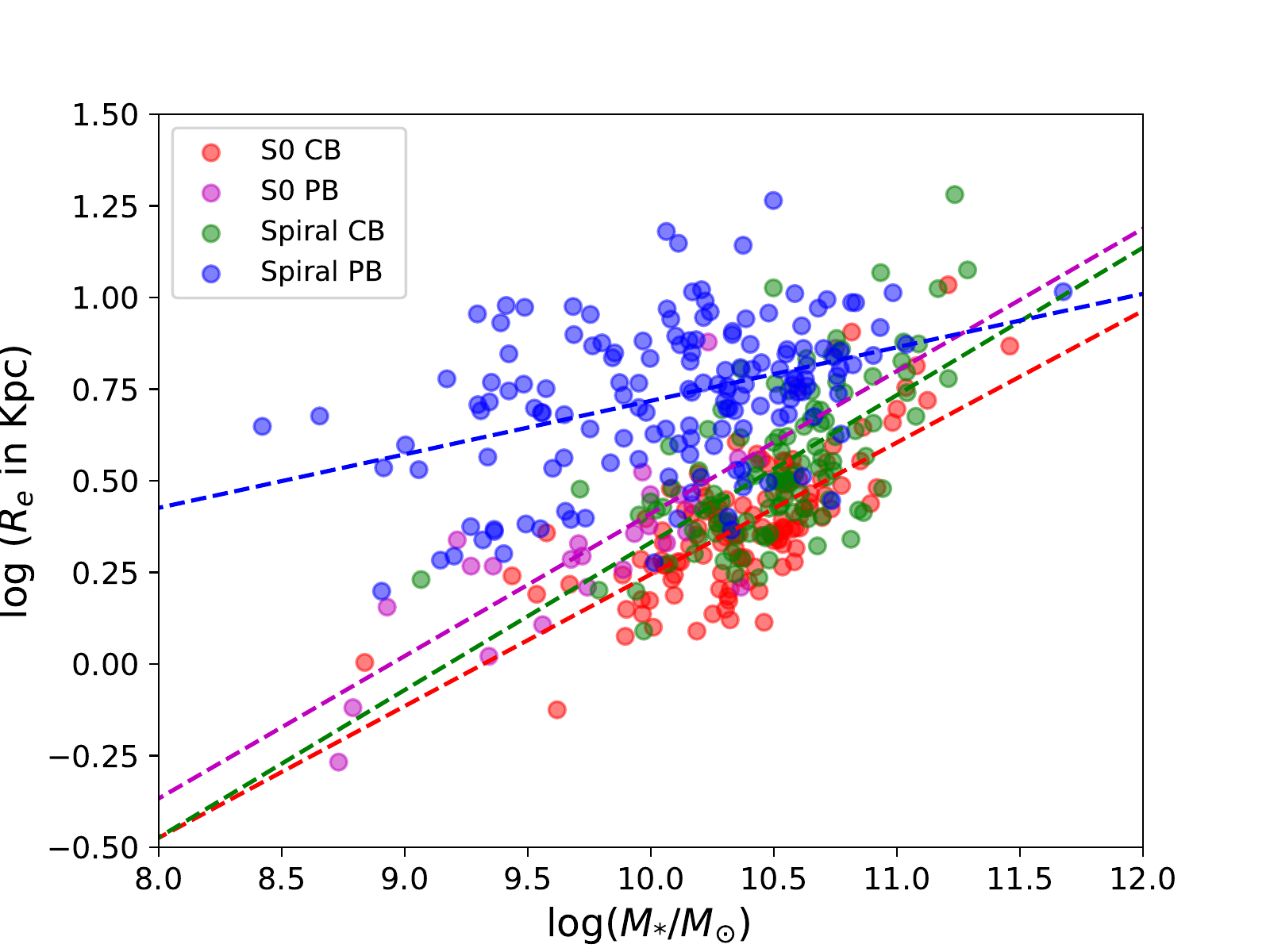}
    
  \end{minipage}
  \hspace{-0.0em}%
  \begin{minipage}[b]{0.49\textwidth}
    \includegraphics[width=\textwidth]{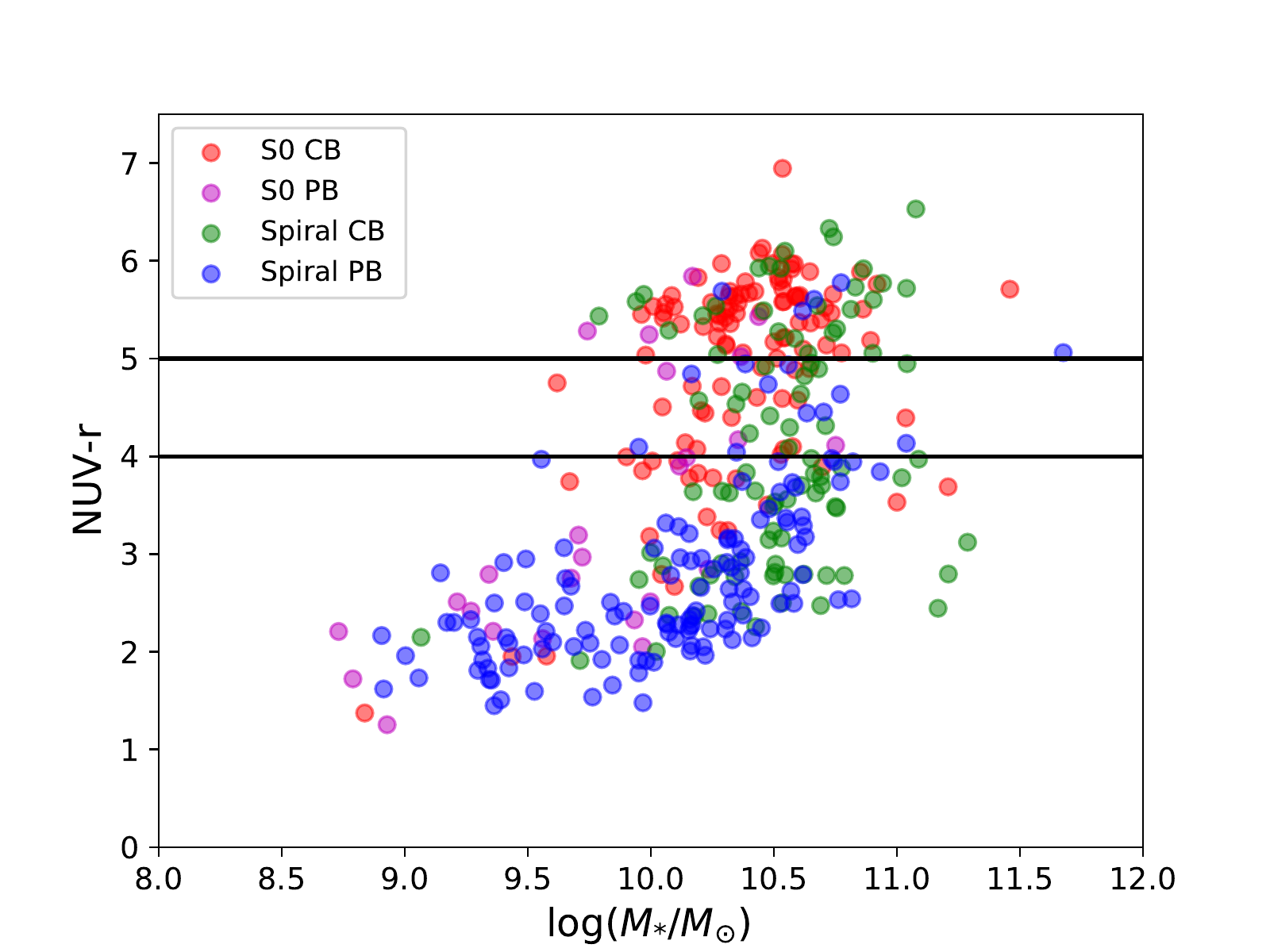}
    
  \end{minipage}
  \caption{The global properties of classical and pseudobulge hosting spirals and S0 galaxies. \textbf{Left} : The size-stellar mass relation for the classical and pseudobulge hosting spirals and S0 galaxies. The galaxy size has been taken as the logarithm of the galaxy semi major axis effective radius ($R_e$). The coloured lines are the best fit lines passing through the different population of objects as denoted by the legend in the plot. The plot shows that, at similar stellar mass, the pseudobulge hosting spirals are bigger in size as compared to classical bulge host spirals. The classical bulge hosting S0 galaxies have a similar mass-size relation as the classical bulge hosting spirals. \textbf{Right}: The NUV-r colour - stellar mass diagram for the spirals and S0s in our sample. The two horizontal lines at NUV = 5 and NUV = 4 mark the boundary of the green valley region \citep{Salim2014} which separates passive red sequence lying above the green valley from the star forming galaxy sequence which lies below this region. The plot shows that pseudobulge hosting spirals are mainly star forming. Classical bulge hosting spirals mainly populate the passive sequence and green valley region just like the classical bulge hosting S0 galaxies. The distribution of galaxies on both panels suggests that classical bulge hosting spirals are the progenitors of classical bulge hosting S0 galaxies.}
\label{fig:both}
\end{figure*} 

We first try to see if there is any difference in properties of the
same bulge type hosted by spirals and S0 galaxies by comparing them
with respect to their kinematics and stellar populations. We have used
the velocity dispersion($\sigma$) and the 4000 {\AA} spectral break index
($D_n(4000)$) to study the kinematics and the stellar population of
the bulge respectively.  The break in the galaxy optical spectrum at 4000
{\AA} arises due to accumulation of stellar absorption lines (of
mainly metals) in the atmosphere of old stars and lack of hot young
stars. This break, quantified by the $D_n(4000)$ index, is larger for
the galaxies having older stellar populations. The $D_n(4000)$ index
is a reliable indicator of the mean age of galaxy stellar populations and
value of $D_n(4000)$ $\sim 1.3$ and $D_n(4000)$ $\sim 1.8$ represents
a stellar population having light weighted mean stellar ages of $\sim$
1-2 Gyr and $\sim 10$ Gyr respectively \citep{Kauffmann2003}. To carry
out the comparison, we have divided our final sample of spirals and S0
galaxies in four mass bins with divisions log$M_*/M_{\odot}$=
[8.5,9.4,10.0,10.6,11.5] and have plotted them on $D_n(4000)-\sigma$
plane for each stellar mass bin as shown in Figure
\ref{fig:Dsigma}. The plots have been divided into four quadrants by
two lines given by $D_n(4000)$=1.5 and $\sigma$=130 km/s. The line at
$D_n(4000)=1.5$ has been used to divide bulges having a young (with
$D_n(4000)<1.5$ ) and an old ($D_n(4000)\geq1.5$) stellar
population. A value of $D_n(4000)=1.5$ corresponds to a stellar age of
$\sim$ 2Gyr. In recent literature \citep{Zahid2017}, this value has
been used to select old and passive galaxies. The line at $\sigma=130$
km/s has been put as a consistency check on our bulge classification
and to identify spurious pseudobulges. \cite{Fisher2016} have
suggested that if a bulge is found to have a central velocity
dispersion greater than 130 km/s , then it is most likely to be a
classical bulge. Therefore, any pseudobulge, in our sample, if found
to have $\sigma>130$ km/s has a high chance of being a misclassified
pseudobulge.

There are a number of interesting trends which can be seen in this
figure. The first thing to notice is that as we go from a lower to
higher stellar mass bins, the bulges become older and dynamically
hot. But in all of the stellar mass bins the velocity dispersion of
pseudobulges remains less than that of classical bulges, on average,
irrespective of the galaxy morphology. One can also notice that there
is sufficient overlap in parameter space between classical bulges
hosted in spiral and S0 galaxies, and majority of classical bulges
tend to have velocity dispersion greater than 100 km/s. Figure
\ref{fig:Dsigma} thus indicates that the classical bulges hosted in
spirals and S0 are not different but are similar in their kinematics
and stellar population properties.\par

After comparing the properties of bulges, we now compare the global
properties of galaxies in our final sample using the galaxy size-mass
relation. In the left panel of Figure \ref{fig:both} we have plotted
the effective radius ($R_e$) versus total stellar mass for all classical and
pseudobulge host spiral and S0 galaxies in our final sample. The
galaxy semimajor axis effective radius ($R_e$) has been taken from
\cite{Simard2011} and the stellar mass estimates are taken from \cite{Kauffmann2003}. The lines passing through each population on this size-mass plot are best fitted straight lines for each population. The plot tells us that pseudobulge hosting spiral galaxies have bigger
sizes as compared to classical bulge hosting spirals at similar
masses. One can also simultaneously see the similarity of mass size
relation between spiral and S0 galaxies which host a classical
bulge.\par

\section{Discussion}
\label{sec:discussion}

We now discuss the probable reason for the observed mismatch of bulge fraction seen in spirals and S0s in the light of results presented in the previous section. We know that morphological transformation of spirals into S0 galaxies can happen through a number of processes. These include the processes driven by interaction with other galaxies, eg. major and minor mergers, tidal harassment etc., or morphological transformation via disc fading due to the shut down of star formation by environmental processes such as ram pressure stripping \citep{Gunn1972} of disc gas or quenching due to starvation \citep{Larson1980}, or by some internal processes like supernova and AGN feedback \citep{Cox2006, Croton2006} or by making the disc stable against star formation due to a massive bulge\citep{Martig2009}. All these processes can change the properties of progenitor spirals in some way or another and hence, the interpretation of the difference seen in bulge fraction in the two morphological types gets linked to the formation scenario of S0 galaxies. As a next step, we discuss the possible ways in which the observed mismatch in classical bulge fraction seen in spirals and S0s can be explained and, how the individual process of morphological transformation fits into the overall picture. \par

If S0 galaxies are transformed spirals then one can think of two possible reasons which can explain the observed mismatch in classical bulge fraction seen in S0 and their supposed progenitors, the spiral galaxies. The first possibility is that the process which transforms spirals into S0 galaxies also changes the bulge type. One mechanism to do so is through galaxy mergers. A major merger of a pseudobulge host spiral with a similar galaxy can produce a classical bulge hosting S0 galaxy and thus increase the classical bulge fraction in the remnant population\citep{Kormendy2016,Tapia2017}. But the extent to which this mechanism can be evoked to explain the observed difference in classical bulge fraction is doubtful due to the reasons mentioned below.\par 

In their study, \cite{Sanjuan2009} have found that only 8\% of present massive disc galaxies (having stellar mass $>10^{10} M_{\odot}$) might have undergone a disc-disc major merger since since z$\sim$1, implying that disc-disc major mergers are not very frequent for the past $\sim$7 Gyrs. This value of merger fraction does not seem to be high enough for significantly changing the classical bulge fraction in the population of merger remnants. Moreover, major mergers of spiral galaxy do not always produce classical bulges and S0 galaxies. In addition to producing elliptical galaxy like remnants via major mergers, recent simulations have also formed a spiral with late type morphology hosting a pseudobulge in major mergers of spiral galaxies \citep{Sauvaget2018}. Finally, a major or a series of minor merger of a pseudobulge hosting spiral galaxy with other galaxies will produce a remnant which is bigger in size than the pseudobulge hosting spirals. If majority of classical bulge hosting S0s are produced by this process, one would have them lying above the pseudobulge hosting spirals in size-mass plane which is not seen in the size-mass relation as shown in the left panel of Figure \ref{fig:both}. Therefore, it seems unlikely that major merger is a dominant reason for the observed mismatch seen in classical bulge fraction of spirals and S0 galaxies.\par

The other possibility which can explain the observed mismatch of classical bulge fraction is that of preferential transformation of classical bulge hosting spirals to S0 morphology and thus increasing the observed classical bulge fraction as compared to the spiral progenitors. The results that we have presented till now, do not oppose the possibility of preferential conversion of classical bulge host spirals to S0 population without changing the bulge type. A spiral galaxy can gravitationally interact with neighbouring galaxies via fly-by interactions and the global tidal field of the environment which can lead to disappearance of spiral arms \citep{Moore1999,Bekki2011}. There are simulations which show that the spirals having higher central surface brightness are more prone to loosing spiral arms due to tidal interactions as compared to the spirals with low central surface brightness \citep{Moore1999}. Usually the surface brightness of a typical classical bulge is higher than a typical pseudobulge \citep{Gadotti2009} and hence there exists a possibility that spirals hosting classical bulges are more likely to transform into an S0 galaxy. However, these simulation have been performed in cluster environment where galaxy-galaxy gravitational interactions are numerous and tidal fields are strong. The galaxies in our sample are selected from a less dense environment. We have checked the available group membership information for the galaxies in our sample from \cite{Yang2007}. Out of total 417 galaxies in our final sample we have group membership information for 393 objects.  We find that out of these 393 galaxies, 296 of them are isolated galaxies and almost all other galaxies reside in groups having 4 or less members. Therefore, it is unlikely that the galaxies in our sample are experiencing strong tidal interaction which is the characteristic of dense cluster environment. For this reason, we believe that the mechanism of tidal interaction due to environment is not driving the discrepancy in classical bulge fraction by selectively converting classical bulge hosting spirals to S0 morphology. \par

  We now consider the next process which can convert spirals to
  S0 galaxies without altering the bulge type.  A process which simply
  shuts down the star formation in the disc of spiral galaxy can lead
  to fading of spiral arms and result in formation of a S0 galaxy
  without significantly altering bulge kinematics and global
  properties of the galaxy such as size and mass. Removal of gas from the
  disc or stopping the fresh infall of gas from the surrounding can
  lead to disappearance of spirals arms \citep{Bekki2002} and make the
  galaxy disc passive \citep{Bekki2009}. Therefore, if pseudobulge
  hosting spirals remain star forming while majority of classical
  bulge host spirals are undergoing star formation shut down then the
  resulting population of S0 galaxies will be dominated by classical
  bulge hosts. The population of S0 galaxies created in this manner
  will also have similar bulge properties and will lie in the same
  size-mass parameter space as classical bulge host spirals as is seen in
  the results presented in the previous section. \par

  We tried to investigate this possibility by studying the star
  formation property of classical and pseudobulge hosting spirals and
  S0 galaxies on NUV-r color-mass diagram which is shown in the right
  panel of Figure \ref{fig:both}. The NUV magnitudes for the galaxies
  in our sample were obtained from the Reference Catalog of Spectral
  Energy Distributions (RCSED) \citep{Chilingarian2017} which is a
  value-added catalogue of 800,299 spectroscopically targeted SDSS
  galaxies in 11 ultraviolet,optical, and near-infrared bands. We
  obtained the $r$ band model magnitude from the SDSS DR 13
  database. In the $NUV-r$ color mass diagram, the two horizontal lines
  at $NUV = 5$ and $NUV = 4$ mark the boundary of the green valley
  region \citep{Salim2014}. The galaxies lying above the green valley
  in the $NUV-r$ color-mass diagram are in the non star forming (quenched)
  sequence while galaxies lying below the green valley are star
  forming. One can notice from this plot that most of spiral galaxies
  seen in quenched sequence are dominated by classical bulge hosting
  spirals. On the other hand, most of the pseudobulge hosting spirals
  are star forming. \par

  The above result supports our hypothesis that the observed
  mismatch between classical bulge fraction in spirals and S0s is due
  to the fact that spirals which host a classical bulge are
  preferentially getting converted into S0 morphology via quenching of
  star formation. The reason for this correlation between bulge type
  and star forming rate and what causes this quenching is not
  clear. However, we do have some clues which can shed some light on
  these questions. Processes such as ram pressure stripping of disc
  gas in cluster environment and virial shock heating of infalling gas
  due to massive dark matter halo can quench a galaxy
  \citep{Gunn1972,Bekki2009,Dekel2006}. Since our final sample spans a
  narrow range of environment, and is dominated by galaxies in the
  field and in small groups having a low mass dark matter halo
  ($<10^{12} M_{\odot}$), we can safely say that quenching due to
  environment, which operates strongly in clusters, cannot be a
  dominant process. The other possibilities include quenching due to
  major mergers or due to some internal processes such as AGN
  feedback. Recently \cite{Weigel2017} have shown that major mergers
  are not a dominant process for galaxy quenching and at any given
  stellar mass the merger quenched galaxies account only for 1-5 \% of
  the quenched population since last 5 Gyr. Therefore, the most
  probable quenching mechanism for the galaxies in our sample seems to
  quenching due to some internal process.\par
  
  Quenching via internal means can happen due to a number of
  mechanisms. One way to make disc passive is by expulsion of disc gas
  through winds driven by intense star formation or supernova
  feedback. But simulations have also shown than quenching due to
  supernova feedback is not enough to sufficiently affect the star
  formation in the disc (\cite{Gabor2010} and references therein). One
  often evokes the feedback from AGN as an additional means to inject
  energy in the surrounding, leading to heating of gas and preventing
  fresh gas infall which leads to quenching. \cite{Bluck2016} have
  studied the correlation between a variety of galactic and
  environmental properties and quenching of star formation. They have
  found that at fixed environment, there is a tight correlation
  between central velocity dispersion and fraction of quenched
  galaxies. They speculate that the reason for this correlation is
  because the central velocity dispersion and black hole mass of a
  galaxy are correlated. A galaxy having higher velocity dispersion
  will host a more massive black hole which then can shut down the
  star formation in the disc via more energetic AGN feedback. We have
  seen in Figure \ref{fig:Dsigma} that classical bulges have higher
  velocity dispersion as compared to the pseudobulges. Therefore, as
  discussed before, there is a possibility that the classical bulge
  host spirals are getting quenched due to black hole feedback. \par

There are other possible internal quenching mechanism like morphological quenching \citep{Martig2009} where bulge dominated galaxies stabilize their discs against star formation due to the presence of a massive bulge. The stability of the disc against star formation leads to quenching of the galaxy with time. This quenching mechanism also predicts that the bulge dominated galaxies form stars less efficiently as compared to the the disc dominated ones. It is known that the bulge to total light ratio is usually found to be higher in classical bulge hosting disc galaxies as compared to the ones hosting pseudobulges \citep{Kormendy2016}. This makes the classical bulge hosting disc galaxies bulge dominant and hence, it is possible that these galaxies are more likely to get quenched morphologically.\par 

All of the mechanism discussed above can lead to disc fading in classical bulge hosting spiral galaxies and can potentially transform them into S0 galaxies. However, we do not have sufficient information to conclusively infer which internal quenching mechanism is actually dominant. The details of internal quenching and how it explains the observed correlation between bulge type and the star formation properties of galaxies is beyond the scope of this work. In future, we would like to study the star formation history of a sample of isolated spirals and S0 galaxies in low density environment by making use of spectroscopic information. Such a study will be valuable for shedding some light on the formation of S0 galaxies in low density environments.

\section{Summary}
\label{sec:sum}

In this work we have tried to understand the possible reason for the
significant difference seen in the classical bulge fraction in the
spiral and S0 galaxy population in the nearby Universe. We have
constructed a sample of 262 spiral and 155 S0 galaxies in a fixed
narrow range of environment and then have classified their bulges in
classical or pseudobulges types using the Kormendy diagram. Dividing
the sample into bins of stellar mass, we have compared the classical
bulge fraction in spirals and S0 galaxies. We have found that, even at
fixed stellar mass, the fraction of galaxies hosting a classical bulge
is higher in S0 galaxies as compared to spirals. If S0 galaxies are
transformed spirals, then this can happen either due to change of
bulge type during this morphological transformation or it can be due
to a process by which classical bulge hosting spirals are more likely
to get converted into S0 morphology as compared to pseudobulge hosting
spirals. Comparing the bulge and global properties of spirals and S0,
we rule out the possibility that the change of bulge type during
morphological transformation being a dominant process and find that
classical bulge hosting spirals are the likely progenitors of
classical bulge hosting S0 galaxies. By studying the star formation
properties of these galaxies we find that majority of pseudobulge
hosting spirals are star forming while those hosting classical bulges
are either in green valley or are in the passive sequence. We
  think that some internal quenching mechanism such as feedback due to
  central supermassive black hole or stability of gas disc against
  star formation due presence of a massive bulge (morphological
  quenching ), shuts down the star formation in these classical bulge
  host spirals and transforms them into classical bulge hosting S0
  galaxies. It is this process that gives rise to the high classical
  bulge fraction, in S0 galaxies. \par

\section*{Acknowledgements}

We thank the anonymous referee for insightful comments
that  have improved  both  the  content and presentation  of
this paper. SB would like to acknowledge South African Astronomical
Observatory (SAAO), Cape Town where part of this work
has been done. YW and SB acknowledge support from a bilateral grant
under the Indo-South Africa Science and Technology Cooperation (PID-102296) funded by the Departments of Science
and Technology (DST) of the Indian and South African Governments.









\bsp	
\label{lastpage}
\end{document}